\documentclass[twoside]{article}

\evensidemargin\oddsidemargin
\advance \baselineskip 0pt plus 0.1pt minus 0.1pt
\input epsf
\usepackage[T2A]{fontenc}

\usepackage{latexsym}
\usepackage{amsbsy}
\usepackage{amsfonts}
\usepackage{amssymb}

\date{}

\begin{document}

\title{
Structures of Order Parameters in Inhomogeneous
Phase States of Strongly Correlated Systems
}
\author{L. S. Isaev, A. P. Protogenov\footnote
{e-mail: alprot@appl.sci-nnov.ru} \\
\\
{\fontsize{10pt}{12pt}\selectfont
{\em
Nizhny Novgorod State University, 603950 Nizhny Novgorod
}\/}\\
{\fontsize{10pt}{12pt}\selectfont
{\em
Institute of Applied Physics of the RAS, 603950 Nizhny Novgorod
}}\/}

\maketitle

\begin{abstract}
The structures of order parameters which determine
the bounds of the phase states in the framework of the $CP^{1}$
Ginzburg-Landau model were considered. Using the formulation
of this model \cite{BFN} in terms of the gauged order parameters
(the unit vector ${\bf n}$,
density $\rho^{2}$ and momentum of particles ${\bf c}$) we found
that some universal properties of phases and field configurations
are determined by the Hopf invariant, $Q$ and its generalizations.
At a sufficiently high level of
doping it was found that beyond the superconducting phase the
charge distributions in the form of loops may be more
preferable than those in the form of stripes.
It was shown that in the phase with its mutual
linking number $L<Q$ the transition
to an inhomogeneous superconducting state with non-zero
total momentum of pairs takes place.
The universal mechanism of the topological
coherence breaking of the superconducting state due to a decrease
of the charge density was discussed.

PACS 74.25.-q, 74.80.-g, 71.10.Hf, 71.10.-w

\end{abstract}

\section*{1. Introduction}

Among the challenging problems of cooperative phenomena in planar systems
near Mott transition, there are those which at the first glance
may not be associated with the appearance of high-temperature
superconducting states in doped antiferromagnetic insulators.
For example, we are interested in the origins of qualitatively similar
cooperative behavior in various compounds and very rich content of
their phase diagram, as well as in origin of the emergence of
inhomogeneous states, the typical for such systems
\cite{L1,L2,H,D,T}. Low-dimensional
structures in the distribution of spin \cite{L1,L2} and
charge \cite{H,D,T}
degrees of freedom exist in the state which precedes the high-temperature
superconducting phase. Keeping in mind this property \cite{PC}
we should choose such a model for describing the mentioned phases which will contain
them as limiting cases. The mean field Ginzburg-Landau theory with the
appropriate choice of the order parameters may be used
for understanding of such a kind general problems.
The key question in this universal approach is the method with
the aid of which the order parameters encode simultaneously
the content and the distribution of charge and spin degrees
of freedom of excitations in various phase states.

The recent progress in solving analogous problems
in the non-Abelian field theory  \cite{FN1} and its development in
condensed matter physics \cite{BFN} has shown
that $CP^{1}$ Ginzburg-Landau model is preferable.
The two-component order parameter of
this model is used for solving the problems of two-gap
superconductivity \cite{B}. In the theory of electro-weak interaction
\cite{Ch}, it has the sense of the Higgs
doublet of the standard model. In this paper we will assume that
the order parameter is a spinor realizing two-dimensional
representation of the braid group which arises due to classifying
the quantum states at permutations of particles
in (2+1)-dimensional systems.
Considering factorization with respect to the centre of this
non-Abelian group we obtain the gauged $CP^{1}$ Ginzburg-Landau model.
We pay attention that the order parameter of this model
is two-dimensional \cite{BFN,AW}. Only in this case we can introduce the unit vector field
which describes the distribution of one-half spin degrees of freedom
in the long-wavelength limit as well as use the Hopf invariant
for classifying ${\bf n}$-field configurations and consider
correctly the phases with different distribution of charge degrees of freedom.
Due to the above mentioned reasons we use the generalized ${\bf n}$-field
model which after the exact mapping of $CP^{1}$
Ginzburg-Landau model \cite{BFN} includes Faddeev term \cite{F}.
Because of the non-Abelian gauge theory origin of this significant part
of the model we hope that the obtained answers are universal
and will give a deeper insight into the discussed problems.

The Hopf invariant describes the degree of linking or knotting
of the filamental manifolds where the field of the unit vector ${\bf n}$
is defined. The study of the behavior of the vortex filament
tangle is a separate problem and attracts attention due to several reasons.
First, at small distances the topological order
associated with the linking exists against the background
of the disorder caused by arbitrary motion of separate parts
of the system of entangled vortex filaments.
Thus, unlike point particles, the properties of the tangle are determined
by the behavior of its fragments in the ultraviolet and infrared limits.
Since the coordinates of the vortex
core are canonically conjugated, the mentioned circumstances are
presented by noncommutativity of these variables depending on the
linking degree. Second, such soft medium as the tangle of linked filaments
is the hot problem in condensed matter physics and beyond,
in particular, in connection with the DNA problem.
After the coupling constants are transformed into the searched functions
the appearance of the double helix as the solution
of the equations of
motion in the soft variant \cite{N}
of the ${\bf n}$-field model is its general property.

In the present paper we consider some properties of
field configurations in $CP^{1}$ Ginzburg-Landau model defined
in the following paragraph.
The main goal of the paper is to find the bounds of free energy
in the superconducting state and in the inhomogeneous phase
with broken antiferromagnetic order, as well as to describe
the properties of the charge density distributions corresponding
to this state. Considering the non-superconducting phase in the soft
version of the model \cite{N} we analyse the value of the contribution
to the free energy of the charge density distributions in the form
of loops and stripes.
Along with the results from brief publications \cite{PV,P,IP}
in the third and
fourth sections, we discuss the properties of the inhomogeneous
superconducting state with nonzero total momentum of pairs and
compare it with the LOFF states \cite{LO,FF}
and with the results from the recently proposed \cite{W}
BCS-like model with two types of particles.
We also pay attention to the dependence of the bounds of phase states
on the generalized (2+1)D Hopf invariant in the case of
$S^2\times S^1\to S^2$ and $S^1\times S^1\times S^1\to S^2$
mapping classes and on the external magnetic field.
In conclusion we discuss some open problems.
In appendix there is the proof of the inequality which determines
the relation between the contributions
to the free energy of ${\bf n}$- and ${\bf c}$-field configuations.

\section*{2. $CP^{1}$ Ginzburg-Landau model}

We will use Ginzburg-Landau model:
\begin{equation}
 F=\int d^{3}x\,\biggl[\sum_{\alpha}\frac{1}{2m}\left|\left(\hbar
           \partial_{k}+i\frac{2e}{c}A_{k}\right)\Psi_{\alpha}\right|^2+
   \sum_{\alpha}\left(-b_{\alpha}|\Psi_{\alpha }|^{2}+\frac{c_
  {\alpha}}{2}|\Psi_{\alpha}|^4\right)+\frac{\bf B^{2}}{8\pi}\biggr]
 \end{equation}
 with a two-component order parameter,
\begin{equation}
  \Psi_\alpha=\sqrt{2m}\,\rho\,\chi_\alpha,\,\,\,\,\chi_\alpha=|\chi_\alpha
  |e^{i\varphi_\alpha}\,,
 \end{equation}
 which satisfies the condition of $|\chi_{1}|^{2}+|\chi_{2}|^{2}=1$.
This constrain of two components $\chi_\alpha $
takes place in the complex projective space $CP^1$,
for which the considered model is defined.
The model (1), (2) with different masses was used before \cite{BFN,B} [1,6]
in the contex of two-gap superconductivity,
as well as in the standard model of the non-Abelian
field theory \cite{FN1,Ch}.
In the present paper we consider the states
in planar systems. Thus we suppose that $\Psi $
has the sense of the order parameter
realizing two-dimensional non-Abelian representations
of the braid group used for classifying quantum states
at permutations of  particles in
systems with two spatial dimensions.
Realizing Abelian projection \cite{PVB}, the vector $A_k$
compensates the local choice of the phase of the
function $\Psi $. The last terms in (1) describe Ginzburg-Landau
potential $V(\Psi_1,\Psi_2)$ and the self-energy of the gauge field.
It has been shown recently \cite{BFN} that there is an exact
mapping of the model (1), (2) into the following
version of the ${\bf n}$-field model:
\begin{eqnarray}
  F = \int d^{3}x\left[\frac{1}{4}\rho^{2}\left(\partial_{k}{\bf n}
         \right)^{2}+\left(\partial_{k}\rho \right)^{2}+\frac{1}{16}\rho^{2}
         {\bf c}^{2}+\left(F_{ik}-H_{ik}\right)^{2}+V(\rho, n_{3})\right].
 \end{eqnarray}
 The free energy in Eq.(3) is defined by the
 scalar -- the density of particles $\rho^2$,
 the field of the unit vector $n^a={\bar \chi}\sigma^{a}\chi$,
 where $\bar\chi=(\chi_{1}^{\ast}, \chi_{2}^{\ast})$,
 $\sigma^{a}$ is Pauli matrix, and the field of the momentum
 ${\bf c}={\bf J}/\rho^{2} = 2({\bf j} - 4{\bf A})$.
 The total current ${\bf J}$ contains paramagnetic part
 ${\bf j}=i[\chi_{1}\nabla\chi_{1}^{\ast}-c.c.+(1 \to 2)]$
 and diamagnetic term $-4{\bf A}$.
 Equation (3) was obtained with the use of the following notations:   $F_{ik}=\partial_{i}c_k-\partial_{k}c_i$,
 $H_{ik}={\bf n}\cdot[\partial_{i}{\bf n}\times\partial_{i}{\bf n}]$,
 and dimensionless units as follows:
 of the length $L=(\xi_1+\xi_2)/2$ with the coherence length
$\xi_{\alpha}=\hbar /\sqrt{2m\,b_{\alpha}}$, the
momentum $\hbar/L$ as the unit of the momentum {\bf c},
the density units $c^2/(512\pi e^2 L^{2})$ of
particles per mass unit (at parametrization of $\Psi_\alpha$
in the form (2)) and the
energy units $\gamma /L$ with  $\gamma=\left(c\hbar /e \right)^2/512\pi$.

In formulation (3) the Ginzburg-Landau functional
depends on gauged order parameters $\rho^2$, ${\bf c}$,
and ${\bf n}$.
They characterize spatial distributions of charge and spin
degrees of freedom with current or without it.
The functions $\chi_\alpha$ determine the orientation
of the unit vector  ${\bf n}$
which describes (in the long-wavelength limit)
the properties of the magnetic order.
Besides, functions $\chi_\alpha$ define the value of the paramagnetic
part of the current. Comparing different forms of
presentation of $CP^{1}$ Ginzburg-Landau model we note
that vortex field configurations $\Psi_\alpha$ in the model (1), (2)
are equivalent to textures of the field ${\bf n}$
in terms of the model (3). We also pay attention
to the fact that the ansatz (2) has the sense  of
factorization of longitudinal $\rho$ and transversal $\chi_\alpha$
degrees of freedom.
In the superconducting state the
composition of spin ${\bf j}$ and charge degrees of freedom is
important, since the
current contains  diamagnetic $U(1)$ gauge component  $-4{\bf A}$.

In the soft variant of the extended model of ${\bf n}$-field (3)
the multipliers of the
first term describe the distributions of spin stiffness
and the square of the inverse length of the density screaning.
It is seen from this example that the competition of the
order parameters  $\rho, {\bf n}$ and ${\bf c}$ may be the origin of the
existence of the phase states
with different ordering of charge and spin degrees of freedom.
We enumerate the limiting cases of the model (3)
in inhomogeneous (${\bf n}\not=const $) situations: \\
1. A state with a broken antiferromagnetic order:
${\bf c}=0$, $\rho=const. $ \\
2. A state with a quasi-one-dimensional density distributions:
${\bf c}=0$, $\rho\not=const. $ \\
3. An inhomogeneous superconducting state:
${\bf c}\not=0$, $\rho=const. $ \\
4. ${\bf c}\not=0$, $\rho\not=const. $ \\
In the case of ${\bf n}=const$ and  ${\bf c}\not=0$, $\rho\not=const$
functional (1) is equivalent to the one-component Ginzburg-Landau
model.

\section*{3. The bounds of the free energy}

1. {\it A phase state with a broken antiferromagnetic order}. --
Let us consider the first case in the mentioned list.
In this limit the free energy equals
\begin{equation}
  F=\int d^{3}x\left[g_1\left(\partial_{k}{\bf n}\right)^{2}
   +g_2\left({\bf n}\cdot\left[\partial_{i}{\bf n}\times
        \partial_{k}{\bf n}\right]\right)^2\right] \, .
 \end{equation}
We supposed that in the considered phase the
constant value $\rho=\rho_0$ may be found
from the minimum of the potential $V$ and introduced
the notation $g_i$ for the coupling constants.
The properties of the model (4) were studied in detail
in Refs. \cite{FN2,GH,BS,HS,VK,KR,Ward}.
The analysis of the dimensionality shows
that the first term in Eq. (4) is proportional to the
characteristic size
$R_{Q}$ of the ${\bf n}$-field configurations, and the second term is
inversely proportional to this scale.
Therefore, the energy (4) has a minimum which is
achieved at $R_{Q}=\sqrt{g_2/g_1}$.
This explains why the second term in Faddeev-Niemi model (4)
allows to avoid Derrik's restriction of the existence
of three-dimensional static configurations
with the finite size. In the infrared limit this term
characterizes the mean degree of noncollinearity
$<0|{\bf S}_1\cdot\left[{\bf S}_2\times {\bf S}_3\right]|0>$ in
the orientation of three
spins which locate the sites of the quadratic placket.

It was shown in Refs. \cite{VK,KR,Ward} that the lower
energy bound in the model (4)
\begin{equation}
  F \geqslant 32\pi^{2}\,|Q|^{3/4}
 \end{equation}
 is determined by the Hopf invariant
 \begin{equation}
  Q=\frac{1}{16\pi^{2}}\int d^{3}x\,\varepsilon_{ikl}a_{i}
    \partial_{k}a_{l}\,.
 \end{equation}
 In this equation the vector $a_i$ denotes the gauge potentrial which
parametrizes the mean degree of the noncollinearity
of the orientation of neighboring spins in the
following way: $H_{ik}={\bf n}\cdot[\partial_{i}
 {\bf n}\times\partial_{k}{\bf n}] \equiv \partial_{i}a_{k}-
 \partial_{k}a_{i}$.

The dependence $F_{min}\sim |Q|^{3/4}$ (5) with the boundary conditions
${\bf n}\to (0,0,1)$ in the space infinity was verified in Refs.
\cite{GH,BS,HS} in simulation of the
configurations of ${\bf n}$-field.
Such a boundary condition means
that the space $\mathbb R^3$ of the ${\bf n}$-field definition effectively
compactizes into a three-dimensional
sphere $\mathbb S^3$. Thus the unit vector ${\bf n}$ accomplishes
the mapping of $\mathbb S^3$-sphere
into the space of the two-dimensional sphere $\mathbb S^2$.
Let the vector ${\bf n}$ be directed to some general
point of a two-dimensional sphere. We will be interested in
the answer to the following question:
what is a pullback of this point in the
space $\mathbb S^3$? Or in other words, what set of points from the
definition field of the vector ${\bf n}(x,y,z)$ contributes
to the point of the target two-dimensional sphere?
Since the space $\mathbb S^3$ is compact and its dimensionality is
greater by unity than that of the sphere $\mathbb S^2$, the pullback
of points on the sphere $\mathbb S^2$ are closed and in general linked
lines on the sphere $\mathbb S^3$.
The Hopf invariant $Q$ (6) describes the degree of
linking or knotting of these lines. It belongs to a set
of integers $\mathbb Z$ to which the considered homotopic group
$\pi_3(\mathbb S^2)=\mathbb Z$ equals.
In particular, for two once-linked circles $Q=1$ equals 1,
for one of the simplest knots (a trefoil) $Q$ equals 6 and etc.
As a result, the ${\bf n}$-field
configurations are divided into classes corresponding
to the values of Hopf
invariant. We should emphasize once again
that linked or knotted configurations may be
numbered by the Hopf index only in the case of the $CP^1$ Ginzburg-Landau
model with its two-component order parameter,
because at при $M>1$ the homotopic group $\pi_3(CP^M)=0$ is trivial
\cite{AW}. \\

2. {\it Quasi-one-dimensional density distributions}. --
Let us consider the states outside the superconducting phase from the second
line of the list of limiting cases, to which the $CP^1$
Ginzburg-Landau model leads.
In this soft version of the model the functional (3) has the form
\begin{equation}
  F=\int d^{3}x\left[\frac{1}{4}\rho^{2}\left(\partial_{k}
    {\bf n}\right)^{2} + \left(\partial_{k}\rho \right)^{2} +
    H_{ik}^{2} - b\rho^2 + \frac{d}{2}\rho^{4}\right]\, .
 \end{equation}
In Eq.(7) the positive constant $b$ correponds to the phase with the broken
antiferromagnetic order.

The state with the broken antiferromagnetic order considered above has a lower
energy than the "soft" state we are interested in now. The latter may be
metastable \cite{PS}. In this section, we will consider just such states and compare their
contribution to the Ginzburg-Landau energy without studying the problems of their
relaxation, the critical sizes of nuclei of different phases and etc., which are
of separate interest.

Under the condition that the electron spin and
charge are transferred from one of four sites of some plaquettes to the
dopant reservoir, the terms with $H_{ik}$ in Eq.(1) characterize (in the infrared
limit) the mean degree of non-collinearity $<0|{\bf S}_1\cdot\left[{\bf S}_2\times {\bf S}_3\right]|0>$ in the orientation of three
spins, which remain in the sites of a quadratic lattice plaquettes. Therewith
the deficit of the charge density $\rho_{h}^{2}$ relates to the density $\rho^{2}$, describing in
(1) the distribution of the exchange integral, by the relation $\rho^{2} + \rho_{h}^{2} = const$.  From the
long-wavelength point of view the distribution of the spin density $\rho^{2}$
in the limited region with the exponential law of decrease at its boundary (for
example, at the distribution in the circle with radius $r_{0}$ with the
exponential decrease at the length $R\ll r_{0}$) will be accompanied by the
quasi-one-dimensional distribution of the charge density $\rho_{h}^{2}$
along the
boundary of this region, i.e. along the ring with thickness $R$ and radius $r_{0}$.
It is seen from here that studying spatial configurations of the density field $\rho^{2}$ in planar systems makes it possible to find the form of one-dimensional distributions of the electric charge density with the aid of the above-mentioned holographic projection.

It has been known for a long time that in such a phase state the distributions
of the charge density $(\partial_k\rho)^2$ have the form of stripes\footnote{We suppose that the stripe characteristic sizes
are essentially greater than the lattice scale. In this case, the use of phenomenological
approach of the Ginzburg-Landau mean field theory is justified.}. Due to the gradient term $(\partial_k\rho)^2$ in (7), quasi-one-dimensional field $\rho$ configurations are really
preferable.
It seems to be almost obvious also that density distribution in the form of
rings give the smallest contribution to the energy. Let us find the
contribution to the free energy (7) from quasi-one-dimensional density
distributions $\rho^2$ in the form of rings and stripes and compare the computation
results with the experimental data. We will choose the following trial
functions for the field $\rho$ configurations in the form of a ring
and a stripe:
\begin{equation}
  \rho=\rho_0\; exp \left[-(r-r_0)^2/2R^2\right]
 \end{equation}
 and
\begin{equation}
 \rho=\rho_0 exp \left[-x^2/2L_{x}^{2}\right] \times \left\{
 \begin{array}{cc}
 1, & |y|\leqslant L_{y} \, , \\
 exp \left[-(|y|-L_y)^2/2L_{x}^{2}\right], & |y|>L_y \, .
 \end{array}\right.
 \end{equation}
Here $\rho_0=\sqrt{b/d}$, $r_0$ is the ring radius, $R$ is its width, $2L_y=2\pi r_0$ is the stripe length, $L_x=R$ is
its width. Since configurations (8) and (9) do not depend on the third
coordinate, we will assume that along it the size is limited by the length
$L_z$ and also that $R\,<\,r_0$. The calculation of energy (7) with the aid of (8) and (9) yields the following results for the contribution to the free energy from the ring $F_r$ (at $R \ll r_{0}$) and the stripe $F_{xy}$:
\begin{equation}
  F_r=\pi\rho^2_0 L_z\frac{\bar{r}_0}{R}\left ( 1+
      \frac{R^2}{\xi^2}\right ) \, ,
 \end{equation}
 \begin{equation}
  F_{xy}=\pi\rho^2_0 L_z\frac{\bar{r}_0}{R}\left ( 1+
         \frac{R^2}{\xi^2}+\frac{R}{\bar{r}_0}+
         \left( n_0-\frac{3}{4}b \right)\frac{R^3}{\bar{r}_0}
         \right) \, .
 \end{equation}
Here $\bar{r}_0=\sqrt{\pi}r_0$, $1/\xi^{2}=2\left [ n_0-
 (1-1/\sqrt{8})b \right]$, $n_0$ is a certain characteristic value of the "multiplier" $\,$ $(\partial_k{\bf n})^2$ in (2),
which is of the order $c_1 R^{-2}$, whereas $b=c_2 R^{-2}\delta T$, where $c_{i} \sim 1$ and $\delta T = (T_{c} -
 T)/T_{c}$. In these equations we
omitted the term $H_{ik}^{2}$ from Eq.(7) since we
consider that it is approximately the same for the both
types of distributions.
The exact equation for $F_r$ (in units $\pi\rho^2_0 L_z$) contains the term
\begin{displaymath}
  \delta F_r = 2\left [ I_{3}(x_0) - x_{0}I_{2}(x_0)\right] + \frac{bR^2}{2}\left[ I_1(x_{0}\sqrt{2})-x_{0}
  \sqrt{2}I_{0}(x_{0}\sqrt{2})\right ]+\frac{R^2}{\lambda^2}
  \left[ I_1(x_0)-x_{0}I_{0}(x_0)\right] \, .
 \end{displaymath}
But it is exponentially small already at $R/r_0\sim 1/4$ and
$R \sim \lambda$ with $1/\lambda^{2}=2(n_{0}-b)$:  $\delta F_r\sim 10^{-7}$.

For the optimal width $R=\xi$ (at $R \ll r_0$)
the difference in free energies $\Delta F=F_{xy}-F_r$ in units
$\pi\rho^2_0 L_z$ has the form: $\Delta F=1+c_1-(3/4)c_2 \delta T $.
One can see from this equation that at $4(1+c_1)/3c_2 < 1$
in the temperature
range $\left [ 1-4(1+c_1)/3c_2 \right]T_{c}\,<\,T\,<\,T_c$ bordering on the critical temperature $T_c$ of the transition to the
state with the spin pseudo-gap, rings are preferable (see Fig. 1).
\begin{figure}
\begin{center}
\epsfxsize=80 mm \epsfbox{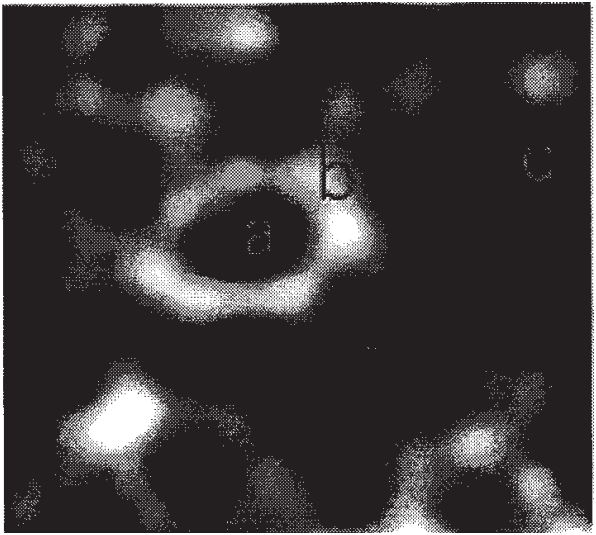}
\end{center}
\caption{
Schematic representation of closed (b) and open (c) quasi-one-dimensional
structures of the charge density (see. \cite{HT}) around antiferromagnetic dielectric
nano-clusters (a).
}
\end{figure}
In the temperature range $T\,<\,T_c\left [ 1-4(1+c_1)/3c_2 \right]$
stripes are the main configurations. As it is known one may approach $T_c$
keeping the temperature constant by increasing the level of doping. The recent
paper \cite{HT} gave the evidence of the existence
of ring-shaped charge structures obtained just
under such experimental conditions. In a certain sense, the tunnel microscope in this
experiment \cite{HT} collects data from a two-dimensional slice of knots \cite{FN2}.

Let us make several remarks concerning the spin density $\rho^2$ distribution in the disk, surrounded by a ring charge distribution. The above-mentioned
statement concerned the case when spin disorder arose only in the region
directly adjacent to the disk edge and, as a result, we had an
antiferromagnetic phase inside. If the antiferromagnetic order is broken everywhere
in the disk, it is necessary to consider the corresponding distribution of the
density $\rho^2$ in the form of a disk in order to compare its contribution
with $F_r$.
When we consider the contribution of the density distributions
$\rho ^{2}$ to the free
energy in the form of a disk, we get a double gain (in comparison with
rings) due to the existence of one edge and have a loss due to the area. The
calculation shows that at small $R/r_0$ the distributions in the form of rings
appear to be more preferable.

Let us consider the dependence of the critical temperature $T_c$
on the level of doping. To do this, we present the relation of $F_r$
to $F_{xy}$ in the following form:
\begin{equation}
  \frac{F_r}{F_{xy}}=\frac{1}{1+B\,R/\bar{r}_{0}} \, ,
 \end{equation}
 where
\begin{displaymath}
B = \frac{1}{2}\biggl [ 1+\frac{1}{2}\frac{n_0 - 3/4\,b}
     {n_0 - (1 - 1/\sqrt{8} )\,b}\biggr]=\frac{3}{4}\;\;
     \frac{n_0-0.68\,b}{n_0-0.65\,b} \, .
 \end{displaymath}
One can see that configurations in the form of stripes are more preferable in
the range $0.65\,<\,n_0/b\,<\,0.68$, where $F_{xy}\,<\,F_{r}$.
Normalizing the density $\rho^2_0$ to the particle number $N$ by
the condition $N=2\pi r_{0}\xi L_{z} m \rho^2_0$, we get the relation between the parameters $n_0$ and $b$,
which we write in the form of $x=b/\sqrt{n_0 - 0.65b}$. Here $x = Nd/(\sqrt{2}m\pi L_z r_0)$. Thus, for the bounds of the
considered above range, where $n_0 \sim b \sim \delta T$, we have $T(x) = T_{c}(1-A\,x^{2})$ with a certain constant $A$.
Therefore, inside the region, belonging to the phase state with the broken
antiferromagnetic order, there is a narrower region, locating between the
parabolas $T(x)$, where the charge structures
have the form of stripes. \\\\

3. {\it The inhomogeneous superconducting state}. --
We consider a superconducting state with finite value of the
total current ${\bf J}$ which exists against the background of a
certain ${\bf n}$-field distribution,
assuming that $\rho = \rho_{0} = const.$
In this case the free energy equals
\begin{eqnarray}
  \lefteqn{F=F_{n}+F_{c}-F_{int}=} \\ \nonumber
  & & \; \int d^{3}x\left[\left(\left(\partial_{k}
         {\bf n}\right)^{2}+H_{ik}^{2}\right)+\left(\frac{1}{4}{\bf c}^{2}+
         F_{ik}^{2}\right)-2F_{ik}H_{ik}\right]\,.
 \end{eqnarray}
The negative sign of the interaction energy $F_{int}$ of ${\bf c}$-
and ${\bf n}$-fields appears because of
diamagnetism of the considered state. As a result the coupling
constant $g_{2}=1$ of
the term $H_{ik}^2$ decreases due to renormalizing
in such a way that  energy of the
superconducting state ${\bf c}\not=0$ is less than the minimum value in the
inequality (5).
To find the exact lower bound of the free energy in the
superconducting state
${\bf c}\not=0$, we will use the following auxiliary inequality
\begin{equation}
  F_{n}^{5/6}\,F_{c}^{1/2} \geqslant (32\pi^{2})^{4/3}\,|L|
  \, \, ,
 \end{equation}
where the invariant
\begin{equation}
  L=\frac{1}{16\pi^{2}}\int d^{3}x\,\varepsilon_{ikl}c_{i}
    \partial_{k}a_{l}
 \end{equation}
determines the degree of the mutual linking \cite{AK,M} of the current lines and the
lines of the magnetic field
${\bf H}=\left[\nabla\times {\bf a}\right]$. Like $Q$ it
is the integral of motion
\cite{M,ZK} in the considered barotropic state.
The proof of inequality (14) \cite{PV}
is given in the appendix of the paper.

Linking indices, which characterize the correlations of spin and charge degrees
of freedom, form the following matrix:
\begin{equation}
 K_{\alpha \beta} =
 \frac{1}{16\pi^{2}}\int d^{3}x \, \varepsilon_{ikl}a_{i}^{\alpha}\partial_{k}a_{l}^{\beta} =
 \left( \begin{array}{cc}
 Q & L^{\prime}  \\
 L & Q^{\prime}
 \end{array} \right)  \, .
 \end{equation}
In this symmetric matrix ($L=L^{\prime}$) with $a_i^1 \equiv a_i$ and
$a_i^2 \equiv c_i$ the integral might be determined also by the
asymptotic linking number \cite{AK}. Let us pay attention
to a circumstance which will
be important below. Being normalized to the charge density,
unlike the unit vector ${\bf n}$, the vector of the momentum
${\bf c}={\bf J}/\rho^2$
belongs to the noncompact manifold. Because of this
the Hopf numbers, defined with the aid of it in (16) are not
integers in general: $(L, Q^{\prime})\not \in \mathbb Z$.
In the superconducting state, where Abelian $U(1)$ gauge symmetry
is broken and the charge is not conserved, the numbers $L$ and
$Q^{\prime}$ play the
role of continuous interpolation parameters,
which unite the considered
compressed and incompressed ($K_{\alpha\beta} \in \mathbb Z$) phases.
From this point of view,
the superconducting states with
$K_{\alpha\beta} \in \mathbb Z$ and
$K_{\alpha\beta} \not \in \mathbb Z$ belong to one and the same class
of universality \cite{RG}.

To find the lower bound of the functional (13), alongside with Eq.(14) we
will use Schwarz-Cauchy-Bunyakovski inequality:
  \begin{equation}
  F_{int}\leqslant 2\| F_{ik} \|_2\cdot \| H_{ik} \|_2 \;
  \leqslant 2F_{c}^{1/2}F_{n}^{1/2} \, .
 \end{equation}
Here $\| F_{ik} \|_{2} \equiv \left[\int d^{3}x F_{ik}^2\right ]^
 {1/2}$. We note that the equality in the r.h.s. of Eq.(17) is
achieved in
the ultraviolet limit, when the size of linked vortex configurations
is small enough.
Substituting the boundary value $F_{int}$ in (13) we get
\begin{equation}
  F\geqslant F_{min}=(F_n^{1/2}-F_c^{1/2})^2.
 \end{equation}

The Hopf configuration with $Q=1$, for which the lower limit takes
place in Eq.(5),
presents two linked rings with radius $R$ and
$(F_{n})_{min} = 2\pi^{2}R^{3}(8/R^{2} + 8/R^{4})|_{R=1} =
32\pi^{2}$. We will assume that also in
our case ${\bf c}\not=0$ there are configurations for which the
equality in Eq.(14) is valid.
Let us emphasize an important circumstance, which we will discuss more
thoroughly in the
next section. For small values of $\rho$ and, therefore, for great values of the
field ${\bf c}$ (since all terms in (13) are of the same order) we encounter the
instability of linked configurations with respect to small perturbations.
This leads to the restriction of values $F_c$ from above.
Keeping in mind this
remark and using in Eq. (18) for $F_c$
of the lower bound $F_{c}^{1/2}=(32\pi^{2})^{4/3}\,F_{n}^{-5/6}\,|L|$
from Eq.(14) and $F_{n}=32\pi^{2}\,|Q|^{3/4}$,
we get for the states with $Q\not=0$, that
\begin{equation}
   F\geqslant 32\pi^{2}\,|Q|^{3/4}\,(1-|L|/|Q|)^{2}\,\,.
 \end{equation}

One can see from Eq.(19) that for all numbers $L < Q$
the energy of the ground state
is less than in the model (4), for which inequality (5) is valid.
The origin of the energy decrease may be understood by comparing
the values of different terms in Eq.(13).
Even under the conditions of a
considerable paramagnetic contribution ${\bf j}$ to the current, the diamagnetic
interaction in the superconducting state for all classes of states
with $L<Q$
reduces in (13) its own energy $F_c$ of the current and a part of
the energy $F_n$
associated with dynamics of ${\bf n}$-field.
In the considered state the total momentum
of superconducting pairs ${\bf c}$ does not equal zero.
In this respect the
inhomogeneous state with current is analogous \cite{AL}
to the state proposed in Refs. \cite{LO,FF}.

\section*{4. The properties of phase states}

The phase state with the broken antiferromagnetic order at
$(\partial_{k}\rho)^2\not=0$ is a
background at which the transition to the inhomogeneous superconducting
phase with $F_{ik}\not=0$ occurs.
It is convenient to discuss the characteristics of this transition at
the density $\rho^2$ change beginning from the superconducting state.
In this phase the constant value of the charge density,
due to the breaking of the gauge invariance $U(1)$,
plays the role of the tuning parameter of the system.

Let the parameter $\rho_0$ change in some range.
Since all terms in Eq.(3) are of the same order, as $\rho_0$
increases, the momentum ${\bf c}$ and, consequently, the index of
the mutual linking $L$ decrease.
In this case at sufficiently small $L$ the smallest
superconducting gap decreases with increase of $Q$ against the
background of a great value of $32\pi^2|Q|^{3/4}$
of the spin pseudo-gap.

When $\rho_0$ reduces, the following effect takes place.
Being proportional to $g_1^{-1/2}\sim\rho_0^{-1}$,
the radius $\mathcal{R}$ of compactification
$\mathbb R^3\to \mathbb S^3$ grows till it exceeds some critical
value $\mathcal{R}_{cr}$.
At $\mathcal{R}\,
 >\,\mathcal{R}_{cr}$ the Hopf mapping is not stable \cite{Ward}
relative to small perturbations of linked
vortex field configurations. As a result, the symmetry $U(2)$
which is associated with identical Hopf mapping appears
to be spontaneously broken.
This means that the topological configurations of
field ${\bf n}$ and ${\bf c}$, instead of being
spread out over the whole space $\mathbb S^3$, localize around a particular point (the base
point of the stereographic projection $\mathbb R^3\to\mathbb S^3$)
and collapse to localized structures \cite{Ward}.
We can see that there is an optimal value of $\rho_0$ and,
consequently, the values
of the characteristic momentum ${\bf c}$
and the relation $|L|/|Q|$, for which there arises
the greatest gain at the transition to the superconducting state.

Till now the vector ${\bf A}$ has characterized the degrees of freedom,
associated with the internal charge gauge symmetry $U(1)$.
If the external electromagnetic field is
applied, the vector ${\bf A}$ equals the sum of internal and external gauge potentials.
In the external magnetic field, due to diamagnetism of the superconducting
state, the momentum  ${\bf c}$ decreases.
Like in the case of $\rho_0$ increasing,
this leads to suppressing the superconducting gap. Playing the
role of the smooth tuning parameter, the external magnetic field
determines the boundary conditions of the problem.
As a result, the answer to the question
of completeness of Meissner screening depends on the results of the
competition of contributions from a paramagnetic (spin) ${\bf j}$
and diamagnetic (charge) $-4{\bf A}$ parts of the total current
${\bf J}$.

Similar to energy distribution in the fractional quantum Hall effect
with the filling factor $\nu=p/q$ and $p,q \in \mathbb Z$ the energy
gain in Eq.(19) depends on the
relation $|L|/|Q|$.
The Hopf invariant $Q \in \mathbb Z$ numbers vacuums \cite{BW} and
is equivalent to the degree of degeneracity $q$
of the ground state. The index $L$  plays the role of the filling
degree $p$ of incomressible charged fluid state
in the fractional quantum Hall effect.
From this point of view the multiplier $(1-|L|/|Q|)$ in Eq.(19) is equivalent
to the filling factor $1-\nu$ for holes.
The distinction of our system from the states in fractional quantum
Hall effect is that the superconducting state is compressible and
here (as it was mentioned)
the effective number $L$ of charge degrees of freedom is not an integer in the
general case.
The configurations of fields ${\bf n}$ and  ${\bf c}={\bf a}$
with the integers $L=Q$, satisfying the relation of self-duality $F_n=F_c$,
correspond to the minimum value of free energy.
In this limit  $K_{\alpha \beta}$
is proportional to the matrix $\left( \begin{array}{cc}
1 & 1  \\
1 & 1
\end{array} \right)$, which was
used in \cite{Wen} to describe the topological order
in the theory of fractional quantum
Hall effect with the filling factor $1-\nu$ at $\nu = 1/2$.

The boundary conditions which determine the momentum ${\bf c}$
and the topological invariants $L,\,Q$ depend not only on
the values of the tuning
parameter $\rho_0$ of the model and the external magnetic field.
Their physical sense and value depend also on the dimensionality of the
manifold for which the model is defined. In the $(3+0)$-dimensional case of the free
energy (3) the Hopf invariant (6) is
analogous to the Chern-Simons action
$(k/4\pi)\int\,dt\,d^{2}x\,\varepsilon_{\mu\nu\lambda}a_\mu
 \partial_\nu a_\lambda$.
This term in the action of $(2+1)$-dimensional systems describes
the dependence of the nonlinear modes' contribution to the free
energy on the statistical parameter $k$. The coefficient $k$ in
the Chern-Simons action has the geometrical sense of the braiding
number of the excitation world lines. In particular, when
semi-fermion excitations (semions) permutate and return to the
initial positions on the plane, the world lines braid twice and
$k=2$. Therewith, the statistical correlations of nonlinear modes
have the character of attraction and for the values $k\sim 2$
their greatest contribution to the energy is of the order of
several percents \cite{APV}. For the energy scale (0.1 -- 1) eV
this gives several tens or hundreds of degrees. Taking into account
the relation between the dimensionality of the systems at their
dynamic and statistical descriptions, we note that the
$(2+1)$-dimensional case $k=2$ with the open ends of excitation
world lines is equivalent (after identification of the ends) to
the compact statistical $(3+0)$-dimensional example of the Hopf
linking with $Q=1$.

The $(3+0)$-dimensional and $(2+1)$-dimensional situations
differ by the topology of the regions of the field definition
${\bf n}$ \cite{V,MM,A}.
When the system is periodic in one of the space variables and
also when calculating the partition function in planar systems,
one of three coordinates -- Matsubara variable -- is a periodic variable.
This means that instead of the sphere $\mathbb S^3$,
we deal with the topology
of a three-dimensional torus $T^{2\times 1}=S^2 \times S^1$
or $T^3=S^1\times S^1\times S^1$
and with the corresponding mapping classes.
The content of Hopf invariant in this case appears to be
richer \cite{V,MM,A}.
For a three-dimensional torus $T^3$
the Hopf invariant is defined modulo $2q$,
where  $q$ is the greatest common divisor of the numbers
$\left \{ q_1, q_2, q_3\right \}\in\mathbb Z$.
Here $q_i$ is the
degree of mapping $T^2 \to S^2$, where $T^2$ is the section
of $T^3$ with the fixed $i$-th coordinate.
Four integral numbers $\left \{ q_i, Q \right\}$,
where $Q$ is defined modulo value $2q$,
give us the complete homotopic classification of mappings
$T^3 \to S^2$ with
$\pi_{1}[Map_{q}(S^2\to S^2)]=\mathbb Z_{2q}$ and a
fixed degree $q$ \cite{V,MM,A}.
The geometrical meaning of this modified Hopf
invariant (an integer from the range
$\left \{ 0, 2q-1 \right\}$) is the same.
It is a linking index of the preimages of two generic points
in $T^3\to S^2$.
The cases $T^{2\times 1}$ and $T^3$ are characterized
physically by different
boundary conditions. The boundary conditions change if
an angular velocity of the rotation of the neutral
superfluid phase in  $He^3$ increases \cite{V,MM}
or an external magnetic field in our
charged system grows. Restricting the Hopf invariant change, the
transition $T^{2\times 1}\to T^3$
promotes the appearence of the incompressible
phase.

\section*{5. Conclusion}

One can see from the given analysis that the gain of the free energy at the
transition to the superconducting state with ${\bf c}\not=0$
arises when there is a
coherent phase associated with the spin degrees of freedom. This phase is
characterised by a pseudo-gap (5) and a topological order associated with
linking. If the density $\rho_{0}^{2}$ is rather great,
the momentum ${\bf c}$ is small and the
transition to the superconducting state is not preferable.
According to our classification, this second state is characterized
by changing values of the order parameters $\rho$ and ${\bf n}$.
The energy loss
due to the term $(\partial_k\rho)^2$ may be reduced because
of the development of
one-dimensional structures.
Whether these one-dimensional charge structures will be open,
forming stripes, or closed almost one-dimensional structures in the form of rings,
depends on the parameters of the potential $V(\rho, n_3)$.
In the phase  $n_3=const$ with neutral spin currents the answer will depend on the value
and the sign of the multiplier  $b(n_{3})$ in the potential
$V(\rho, n_{3})=-b\rho^2+\frac{d}{2}\rho^4$
If $b\,>\,0$, then far from  $T_c$ charge structures with open ends are
preferable \cite{N}, and in the case $T\to T_c$ we should prefer
rings.
The first experiments that verified the existence of charge
structures in the form of rings in the underdoped phase of planar systems were
described in paper \cite{HT}.

The superconducting phase occupies on the "temperature-charge density" $\;$ phase diagram
only a part of the region, belonging
to the phase with the broken antiferromagnetic order.
The bounds of its existence on the phase diagram,
associated with the characteristic values of the density
$\rho_{0}^2$,
are determined at great $\rho_0$ by the inequality (14) \cite{PV}
and at small $\rho_0$ these bounds depend on the critical size
of the knot \cite{Ward},
beginning from which instability of the Hopf mapping arises.

Comparing the results of this paper based on consideration of the
local fields with the conclusions following from the BCS-like model
\cite{W} with two species of fermions, we pay attention to the
following qualitative coincidence. One can conclude from Eq.(3)
that the parameter $\rho_0$ determines the value of the coupling
constant. Therefore, the appearance of the solutions (different
from the standard BCS model) for the superconducting gap in the
paper \cite{W} with the finite value of the coupling constant is
analogous to the existence of the threshold for small values of
$\rho_0$ in this paper.

As distinct from the model \cite{W},
the states considered in the present paper are
significantly inhomogeneous.
The analysis of the state $F_{ik}\not=0,\,\rho\not=const$
is an open problem at present.
Here we mention only that the superconducting
current with the amplitude ${\bf c}_{0}$, flowing round the rings (8),
gives the
additional term ${\bf c}_{0}^{2}R^{2}$ to the multiplier in Eq.(10).
This explains in particular why the superconducting region on the
"temperature--the level of doping" phase diagram  is shifted
to the line $\delta T(x)=0$ of the transition to the state with the spin pseudo-gap.
Indeed, in this case
$V_{eff}(\rho ,
 n_{3}) = -b_{eff}\rho^{2} + (d/2)\rho^{4}$ with
$b_{eff} = b - (n_{0} + {\bf c}_{0}^{2}) =
(const/R^{2})\delta T$.
Therefore the finite value of the momentum ${\bf c}$ of
superconducting pairs
decreases $\delta T$. Besides, the contribution
to the free energy in the inhomogeneous
state due to $(\partial_{k}\rho)^{2}\not=0$ decreases
the gain in Eq.(19).

The superconductivity in cluster systems may be evidently studied,
basing on the approaches beyond the mean field theory. For
example, we may use the exact Richardson solution and Bethe ansatz
equation \cite{DP}, as well as the methods of the conformal field
theory \cite{GS}. The exact solution of the ground state problem
under the condition of the finite value of the total momentum of
pairs at such an approach is one of the important problems. Since
the conformal nature of the dimensionality $3/4$ \cite{AK} in (5)
and (14) influences the character of the scale which enters into
energy dependent response functions (which is proportional to
$T^{3/4}$ \cite{CPR}), it  should also be studied carefully.

We are grateful to A.G. Abanov, S.A. Brazovsky,
S. Davis, L.D. Faddeev, S.M. Girvin, V.E. Kravtsov, E.A. Kuznetsov,
B. Lake, A.I. Larkin, A.G. Litvak, C. Renner, H. Takagi, V.A. Verbus, G.E.
Volovik, Yu Lu, Y.-S. Wu for advices and useful
discussions. This paper was supported in part by the Grant
of the Russian Foundation for Basic Research No. 01-02-17225.

\section*{Appendix}

The proof of the Eq. (14) uses the following chain of inequalities:
 \begin{eqnarray}
  \lefteqn{|L|<||{\bf c}||_6\,\cdot\,||{\bf H}||_{6/5}\leqslant 6^{1/6}
           ||[\nabla\times{\bf c}]||_2\,\cdot\,||{\bf H}||_{6/5}
           \leqslant} \\\nonumber
  & & \;\,\,6^{1/6}||[\nabla\times{\bf c}]||_2\,\cdot\,||{\bf H}||_{1}^{2/3}
            \,\cdot\,||{\bf H}||_{2}^{1/3}\leqslant \\\nonumber
  & & \;\,\,(32\pi^2)^{-4/3}F_{c}^{1/2}F_{n}^{2/3}F_{n}^{1/6}=(32\pi^2)^{-4/3}
            F_{c}^{1/2}F_{n}^{5/6} \, .\nonumber
 \end{eqnarray}
Here $||{\bf H}||_{p}\equiv \left( \int\,d^{3}x\,|{\bf H}|^p
 \right)^{1/p}$ At the first and third steps we used H\"older
inequality: $||{\bf f}\,\cdot\,{\bf g}||\leqslant ||{\bf f}||_{p}\,\cdot\,
  ||{\bf g}||_{q}$ with $1/p+1/q = 1$.
Under the condition  $\nabla \cdot{\bf c}=0$ we employs at the
second step  the Ladyzhenskaya inequality \cite{OAL} [40]: $||{\bf
c}||_{6}\leqslant 6^{1/6}
 ||[\nabla\times{\bf c}]||_2$. The fourth step
in the set of inequalities arises after the
comparison of the terms $||[\nabla\times{\bf c}]||$
 and $||{\bf H}||$ with the terms $F_n$ and $F_c$ in
Eq.(13). The last line shows also separate contributions from
${\bf n}$- and ${\bf c}$-parts of
the free energy (13) to the finite result (14). Using a chain of Holder and
Ladyzhenskaya inequalities, analogously one may find that:
\begin{displaymath}
  F_{n}^{1/2}F_{c}^{5/6}\geqslant (16\pi^2)^{4/3}|L^{'}|  \, .
 \end{displaymath}
The coefficient in this inequality differs from (14) due to the
coefficient 1/4
(because of the charge $2e$ of superconducting pairs)
of the first term of the free energy $F_c$ in (13).

\end{document}